\documentclass[aps,nofootinbib,longbibliography,prd,twocolumn]{revtex4-1}
\usepackage{graphicx}
\usepackage{amssymb,amsmath}
\usepackage[colorlinks,citecolor=blue,linkcolor=blue,urlcolor=blue]{hyperref}
\usepackage{mathrsfs}
\usepackage{enumerate}
\usepackage{bbm}
\usepackage{comment}
\usepackage{tikz}
\usetikzlibrary{arrows}
\usepackage{dsfont}

\def\bra#1{\mathinner{\langle{#1}|}}
\def\ket#1{\mathinner{|{#1}\rangle}}
\def\braket#1{\mathinner{\langle{#1}\rangle}}
\def\N{\mathbb{N}}
\def\bint{\{0,1\}^*}
\def\be{\begin{equation}}
\def\ee{\end{equation}}

\newcommand{\tr}{\mathrm{Tr}}
\newcommand{\scup}{\vee}

\newcommand{\port}{\!\!:\!\!}

\newtheorem{definition}{Definition}
\newtheorem{postulate}{Postulate}

\newcommand{\pcom}[1]{{\color{cyan}{PA: #1\ $\longrightarrow$}}}

\newcommand{\VL}[1]{#1}

\input{macro_nodes_inline.tex}
\begin{document}
\title{On quantum superpositions of graphs, no-signalling and covariance}


\author{Pablo Arrighi}
\affiliation{Université Paris-Saclay, Inria, CNRS, LMF, 91190 Gif-sur-Yvette, France and IXXI, Lyon, France}

\author{Marios Christodoulou}
\affiliation{Institute for Quantum Optics and Quantum Information (IQOQI) Vienna, Austrian Academy of Sciences, Vienna Center for Quantum Science and Technology (VCQ), Faculty of Physics, University of Vienna, Austria}

\author{Amélia Durbec}
\affiliation{Université Paris-Saclay, CNRS, LISN, 91190 Gif-sur-Yvette, France}


\begin{abstract}
We provide a mathematically and conceptually robust notion of quantum superpositions of graphs. We argue that, crucially, quantum superpositions of graphs require node names for their correct alignment, which we demonstrate through a no-signalling argument. 
Nevertheless, node names are a fiducial construct, serving a similar purpose to the labelling of points through a choice of coordinates in continuous space. Graph renamings, aka isomorphisms, are understood as a change of coordinates on the graph and correspond to a natively discrete analogue of continuous diffeomorphisms. 
\VL{We postulate renaming invariance as a symmetry principle in discrete topology of similar weight to diffeomorphism invariance in the continuous.}
We explain how to impose renaming invariance at the level of quantum superpositions of graphs, in a way that still allows us to talk about an observable centred at a specific node.


\end{abstract}

\maketitle

\section{Introduction}

This interdisciplinary work draws its motivations and methods from from recent trends in both Computer Science and fundamental Physics. 

{\em The view from Computer Science.} A fundamental question in Computer Science is: `What is a computer?' That is, what features of the natural world are available as resources for computing and how can we capture these resources into a mathematical definition, a `model of computation'? The Turing machine was believed to be the ultimate such model during the entire first half of the previous century. In the 1960's, spatial parallelism came to be recognised as a major additional resource, captured into distributed models of computing: dynamical networks of interacting automata. In the 1990's, it became clear that quantum parallelism is another powerful computational resource. This was again captured into models of quantum computing, e.g. the quantum Turing machine. Therefore, the current `ultimate' answer to the question `What is a computer?' ought to be a distributed model of quantum computing, e.g. along the lines of a dynamical network of interacting quantum automata. This intuitive idea of a network of quantum computers has been otherwise coined as the `quantum internet' \cite{QuantumNetworksKimble,QuantumNetworksCirac,QuantumNetworksBianconi}. However, in most of these proposal the dynamics of the network is classical, when it is natural to consider the network itself could evolve in quantum manner \cite{ArrighiQCGD}. Then, the connectivity/topological structure of the network itself could be found in a superposition, offering a potentially powerful resource e.g. for improved channel capacities as demonstrated in \cite{ChiribellaSecondShannon}. Such graph superpositions may also be used to capture indefinite causal orders \cite{ArrighiAQG}, with applications to algorithmic \cite{CostaComplexity} and communication complexity \cite{OreshkovQuantumOrder}. This raises the question, then: `What is the appropriate mathematical framework to describe graph superpositions?'. Moreover, it is well-known that the computing power of distributed algorithms crucially depends on the availability of a unique identifier at each node \cite{FraigniaudOblivious}. We may indeed consider that these identifiers aka names are fiducial, serving just to describe the network, when really only its topology should matter. The limitations brought by this requirement (cf. `anonymous networks' \cite{AnonymousNetworks}; `identifier-oblivious' algorithms \cite{FraigniaudOblivious0}) are well-studied classically. This in turn raises the question: `How should renaming invariance be imposed in the context of quantum dynamics over graph superpositions?'.

{\em The view from physics.} A striking feature of most approaches to quantizing gravity is that spacetime is allowed to be in superposition of macroscopic (semi)classical geometries. This genuine quantum gravitational effect ought to  take place already when quantum manipulating weakly gravitating, non-relativistic matter \cite{Christodouloupossibilitylaboratoryevidence2019}. 
The past few years have seen a concentrated effort to devise and implement experiments that ought to confirm or invalidate this phenomenon  \cite{BoseSpinEntanglementWitness2017,BoseMASSIVE2018,Christodouloupossibilityexperimentaldetection2018b,MarshmanLocalityEntanglementTableTop2019}. 
Superposition of macroscopic geometries are also the subject of recent intense discussion in the context of indefinite causal structures \cite{ZychBellTheoremTemporal2017,PaunkovicCausalordersquantum2019}, shedding a quantum informational perspective upon quantum gravitational physics. 

Numerous approaches to quantum gravity employ graphs as the topological skeleton on which the theory is built. Notable examples are loop quantum gravity, causal dynamical triangulations, causal sets and tensor models. The description of a superposition of spacetimes would then involve a superposition of graphs, whose definition is usually avoided. Again the questions arise: `What is the appropriate mathematical framework to describe graph superpositions? How should invariance under changes of coordinates be imposed?' The manner in which theses questions are answered, as we will see in this paper, can have profound consequences for the structure of the theory and its predictions.

 Typically in those theories, invariance under changes of coordinates is attempted by either embedding the graphs in a manifold, or by defining the localisation on a graph using auxiliary physical fields, or by working with a state space defined upon equivalence classes of graphs under renamings. We argue that all of these strategies suffer major drawbacks, and propose an alternative strategy tailor conceptually clear and powerful: to directly use node names as natively discrete coordinates, forgo embedding in the continuous, and impose invariance under renamings at the level of observables rather than at the level of states.  We show in particular that directly working at the level of invariant states misses crucial physics, while using auxiliary fields to rectify the situation can introduce instantaneous signalling.

{\em Contributions to common grounds.} We provide a robust notion of quantum superposition of graphs. This can serve as the basis on which to build the state space for quantum circuit paradigms that admit superpositions of circuits, or for quantum gravity theories featuring superpositions of spacetimes. On the Quantum Computing side we import techniques from the paradigms of Quantum Walks and their multi-particle regime the Quantum Cellular Automata \cite{ArrighiOverview}, together with their recent extension to dynamical graphs \cite{ArrighiQCGD}. On the quantum gravity theory front, we mainly draw inspiration from formulations of Loop Quantum Gravity (LQG) \cite{RovelliQuantumGravity2004}, whose kinematical state space is spanned by coloured graphs, called the spin network states.

 The main subtlety lies in the treatment of a central symmetry we identify and formalise : renaming invariance.

\section{Graph superpositions: names matter}

Let us begin by discussing the state space of quantum superpositions of graphs informally. Precise definitions of these notions are given in Section \ref{sec:labelledGraphs}, and dynamics are defined as a (unitary) operator over this state space. The subtlety in the construction lies on how to treat node names, which we now discuss. 

 Consider a Hilbert state $\mathcal{H}_\mathcal{G}$ defined as that generated by a countably infinite orthonormal basis $\mathcal{G}$, where the elements of $\mathcal{G}$ are \emph{graphs}, denoted as $\ket{G},\ket{G'},\ldots \! \in \mathcal{G}$. That is, each graph labels a different unit vector $\ket{G}$, and a generic (pure) state in $\mathcal{H}_\mathcal{G}$ is a superposition of graphs:
$$\ket{\psi} = \alpha_G \ket{G} + \alpha_{G'} \ket{G'} + \ldots$$
with $\alpha_G,\alpha_{G'}, \ldots \in \mathbb{C}$. The inner product on $\mathcal{H}_\mathcal{G}$ is defined by linearity and by
\begin{equation} 
\langle G \vert G' \rangle = \delta_{GG'}.\label{eq:inner}
\end{equation}
where $\delta_{GG'}$ is unit if $G=G'$ and vanishes otherwise.

Now, two graphs $\ket{G}$ and $\ket{G'}$ can differ only by certain names given to their nodes. Anticipating notation, we denote this as $\ket{G'} = R\ket{G}$ where $R$ is a renaming. Clearly, these two graphs are physically equivalent. On this basis, whenever two graphs differ only by a renaming this suggests that we should take 
\begin{equation}
\langle G \vert R G \rangle=1 \ \ ?
\end{equation}
  In fact, as we will see in the next section, it is imperative that we take them as orthogonal, taking 
\begin{equation}
\langle G \vert R G \rangle=0
\end{equation}
\renewcommand{\vec}[1]{#1}

An analogy can be made with plane waves in quantum mechanics and translation invariance. Consider the  plane wave state $ \ket{p,0} = \int e^{i px} dx$ and the same state shifted in position as $ \ket{p,\Delta}= \int e^{i p(x+\Delta)} dx$, with $p$ the wave momentum and $x$ the position variable. In empty space there is no physical sense in which the two plane waves are different: the shift in position is immaterial, as the plane wave homogeneously spreads across space. Yet, the inner product $\langle \vec{p,0} \vert \vec{p,\Delta} \rangle $ need not be taken to be unit. In fact, it is imperative to distinguish mathematically between $\ket{p,0}$ and $\ket{p,\Delta}$ if we wish to do quantum mechanics. For instance, as a particle propagates, its plane waves components typically evolve from $\ket{p,0}$ and $\ket{p,\Delta}$. Thus, whilst $\ket{p,0}$ and $\ket{p,\Delta}$ alone do not hold any physically relevant position information, their relative difference does: it carries physically relevant relative position information. To avoid confusion we emphasize that in this analogy there does exist an ambient spacetime geometry (Newtonian spacetime) with respect to which distances are defined, while in our treatment of quantum graphs we define superpositions of graphs already at the pre-geometrical, topological level, using only the naming of nodes on the graph.

The heart of the matter then lies in the following observation. If we do not further constrain the theory, having taken $\braket{G|G'}=0$ for graphs only differing in their naming, names will in principle be observable. From a Physics point-of-view this is clearly unreasonable. In the continuum, observables that read out the coordinates of a point on the manifold are excluded by the requirement of diffeomorphism invariance, also known as general covariance. This central insight of general relativity ensures that no prediction of the theory depends on the coordinate system in use. Invariance of graphs under renamings should be recognized as a symmetry of similar importance for theories employing graphs rather than manifolds as their underlying topological space. Again from a Computer Science point-of-view this may or may not be reasonable, depending upon the availability of single identifiers, although ultimately, at low levels of abstractions a computer is but a physical device. 

\section{Named graphs versus instantaneous signalling}\label{sec:signalling}


In this section we present our main argument for using graph names rather than an auxiliary physical field in order to designate nodes, and that it is necessary to work at the level of a state space spanned by states corresponding to named graphs, rather than a state space spanned by equivalence classes of graphs under renamings (aka `anonymous graphs'). We may also call the latter `name invariant states' and the former `name variant states'.
\VL{This invariance under renamings is to be understood as a type of gauge invariance of the state space.} 

In particular, we show that  employing node names ensures we do not inadvertently introduce  instantaneous signalling. We demonstrate this point with a simple toy example. 

Consider the state space to be the span of circular graphs having $n$ number of nodes and links. Each of these nodes has a unique name (e.g. $w, x, y, z$) and can be in any of the following internal states (colours): empty, occupied by an $a$--moving particle, occupied by a $b$--moving particle, or, occupied by both. Nodes have ports $:\!\!a$ and $:\!\!b$, upon which the neighbouring nodes are attached. An $a$--moving particle is depicted as a half filled disk on port $a$'s side and similarly for $b$. Thus, each node has the state space $\mathbb{C}^4$. The global Hilbert space is ${\cal H}=\bigotimes_{n \in V_G} \mathbb{C}^4$, where $\bigotimes_{n \in V_G}$ denotes the tensor product over the nodes of graph $G$.

We take the dynamics to be the simplest known quantum walk, the Hadamard quantum walk, or rather its extension to the multi-particle regime aka quantum cellular automata \cite{ArrighiOverview}. It consists in a unitary operator driving a particle on a lattice in steps. Many quantum algorithms can be expressed in this manner, the Hadamard quantum walk in particular has been implemented on a variety of substrates such as an array of beamsplitters traversed by photons \cite{Sciarrino}. Mathematically, evolution is implemented by applying an operator $U=TH$ on the graph state, the alternation of steps $H$ and $T$. 

The step $H$ is the application of the Hadamard gate to the internal state of each node. Formally, $H=\bigotimes_n \textrm{Hadamard}$ with
\begin{align*}
\textrm{Hadamard}=\left(
\begin{array}{cccc}
1& 0& 0& 0\\
0& \frac{1}{\sqrt{2}} &\frac{1}{\sqrt{2}} &0\\
0& \frac{1}{\sqrt{2}} &-\frac{1}{\sqrt{2}} &0\\
0& 0& 0& 1 
\end{array}.
\right)
\end{align*}
Henceforth, we adopt an easily tractable pictorial notation:
\begin{align*}
H \ket{\splittednode{white}{white}} &= \ket{\splittednode{white}{white}}\\
H \ket{\splittednode{white}{black}} &= \frac{1}{\sqrt{2}} \left( \ket{\splittednode{white}{black}} + \ket{\splittednode{black}{white}} \right) \\
H \ket{\splittednode{black}{white}} &= \frac{1}{\sqrt{2}} \left( \ket{\splittednode{white}{black}} - \ket{\splittednode{black}{white}} \right)\\
H \ket{\splittednode{black}{black}} &= \ket{\splittednode{black}{black}}
\end{align*}
Once $H$ is applied, step $T$ moves the `particles' through port $a$ or $b$ to the adjacent node according to their species, all at once. For instance
\begin{eqnarray}
&T \ket{\raisebox{-6pt}{\resizebox{70pt}{20pt}{\tikz{
\draw (-0.5,0) -- (2.5,0);
\node at (0,0) [draw,rotate=90, circle, circle split part fill={white,white}]{};
\node at (0,-0.4) {$u$};
\node at (-0.3,0)[above]{\scriptsize :$a$};
\node at (0.3,0)[above]{\scriptsize :$b$};
\node at (1,0) [draw,rotate=90, circle, circle split part fill={black,white}]{};
\node at (1,-0.4) {$v$};
\node at (0.7,0)[above]{\scriptsize :$a$};
\node at (1.3,0)[above]{\scriptsize :$b$};
\node at (2,0) [draw,rotate=90, circle, circle split part fill={white,white}]{};
\node at (2,-0.4) {$w$};
\node at (1.7,0)[above]{\scriptsize :$a$};
\node at (2.3,0)[above]{\scriptsize :$b$};
}}}} = \ket{\raisebox{-6pt}{\resizebox{70pt}{20pt}{\tikz{
\draw (-0.3,0) -- (2.3,0);
\node at (0,0) [draw,rotate=90, circle, circle split part fill={black,white}]{};
\node at (0,-0.4) {$u$};
\node at (-0.3,0)[above]{\scriptsize :$a$};
\node at (0.3,0)[above]{\scriptsize :$b$};
\node at (1,0) [draw,rotate=90, circle, circle split part fill={white,white}]{};
\node at (1,-0.4) {$v$};
\node at (0.7,0)[above]{\scriptsize :$a$};
\node at (1.3,0)[above]{\scriptsize :$b$};
\node at (2,0) [draw,rotate=90, circle, circle split part fill={white,white}]{};
\node at (2,-0.4) {$w$};
\node at (1.7,0)[above]{\scriptsize :$a$};
\node at (2.3,0)[above]{\scriptsize :$b$};
}}}} \nonumber
\\
&T \ket{\raisebox{-6pt}{\resizebox{70pt}{20pt}{\tikz{
\draw (-0.5,0) -- (2.5,0);
\node at (0,0) [draw,rotate=90, circle, circle split part fill={white,white}]{};
\node at (0,-0.4) {$u$};
\node at (-0.3,0)[above]{\scriptsize :$a$};
\node at (0.3,0)[above]{\scriptsize :$b$};
\node at (1,0) [draw,rotate=90, circle, circle split part fill={white,black}]{};
\node at (1,-0.4) {$v$};
\node at (0.7,0)[above]{\scriptsize :$a$};
\node at (1.3,0)[above]{\scriptsize :$b$};
\node at (2,0) [draw,rotate=90, circle, circle split part fill={white,white}]{};
\node at (2,-0.4) {$w$};
\node at (1.7,0)[above]{\scriptsize :$a$};
\node at (2.3,0)[above]{\scriptsize :$b$};
}}}} = \ket{\raisebox{-6pt}{\resizebox{70pt}{20pt}{\tikz{
\draw (-0.3,0) -- (2.3,0);
\node at (0,0) [draw,rotate=90, circle, circle split part fill={white,white}]{};
\node at (0,-0.4) {$u$};
\node at (-0.3,0)[above]{\scriptsize :$a$};
\node at (0.3,0)[above]{\scriptsize :$b$};
\node at (1,0) [draw,rotate=90, circle, circle split part fill={white,white}]{};
\node at (1,-0.4) {$v$};
\node at (0.7,0)[above]{\scriptsize :$a$};
\node at (1.3,0)[above]{\scriptsize :$b$};
\node at (2,0) [draw,rotate=90, circle, circle split part fill={white,black}]{};
\node at (2,-0.4) {$w$};
\node at (1.7,0)[above]{\scriptsize :$a$};
\node at (2.3,0)[above]{\scriptsize :$b$};
}}}}
\end{eqnarray}
Nothing  in particular happens if particles cross over each other or land on the same node (no collisions). 

\begin{figure}[h!]\centering
\includegraphics[width=\columnwidth]{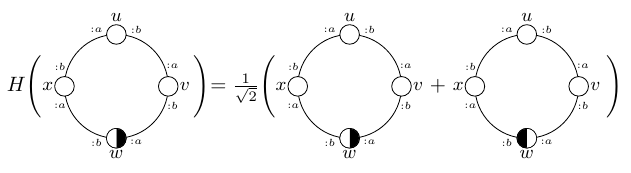}
\includegraphics[width=\columnwidth]{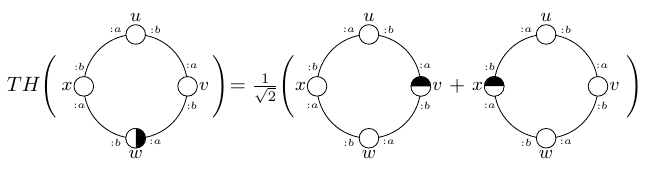}
\includegraphics[width=\columnwidth]{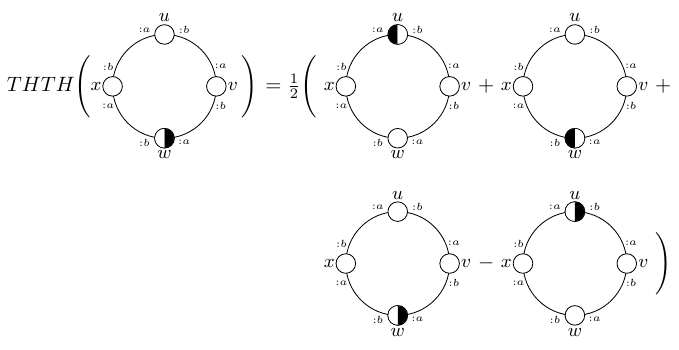}
\caption{\label{fig:UU}Twice the Hadamard quantum walk on circular graphs}
\end{figure}

Now, let us apply $U$ twice on an initial state featuring a single b--moving particle. The computation can be followed pictorially in Fig. \ref{fig:UU}. 
In the final state the node names or `particles position' and their colours are correlated. Note that the first two branches of the superposition and the last two branches of the superposition differ only by a renaming of the nodes ( $u \leftrightarrow w, x \leftrightarrow v$).  The last two branches are those that include a $b$-moving particle and come with an opposite sign. Thus, if names were not present we would be left  with radically different physics, a state that has only $a-moving$ particles, and no entanglement.

\begin{figure}[h!]
\includegraphics[width=\columnwidth]{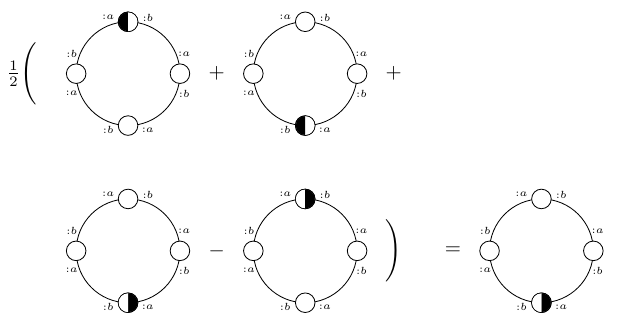}
\caption{\label{fig:UU2}Twice the Hadamard QW on circular abstract graphs}
\end{figure}

To drive this point through we switch to a different Hilbert space ${\cal H}'$ defined as the span of name invariant states or `anonymous graphs'. Technically, anonymous graphs can be defined as equivalence classes of graphs up to arbitrary renamings, see section \ref{sec:labelledGraphs}. A pictorial depiction of this equivalence relation $\sim$ is as below: 
\begin{center}
\newcommand{\glue}{\!\!\!\textrm{\raisebox{0.2684pt}{---}}\!\!\!\!}
\scalebox{0.75}{
\tikz {
\tikzset{
    >=stealth',
    punkt/.style={
           rectangle,
           rounded corners,
           draw=black, very thick,
           text width=6.5em,
           minimum height=2em,
           text centered},
    pil/.style={
           ->,
           thick,
           shorten <=3pt,
           shorten >=3pt,}
}

\node[] at(0,0){\lineoffour{white}{white}{white}{black}{white}{white}{white}{white}};
\node[] at(3,0){$\neq$};
\node[] at(6,0){\lineoffour{white}{white}{white}{white}{white}{black}{white}{white}};
\node[] at(0,-2){\lineoffourmodulo{white}{white}{white}{black}{white}{white}{white}{white}};
\node[] at(3,-2.1){$ \sim $};
\node[] at(6,-2){\lineoffourmodulo{white}{white}{white}{white}{white}{black}{white}{white}};

\draw[pil,>=latex] (0,-0.3)--(0,-1.7);
\draw[pil,>=latex](6,-0.3)--(6,-1.7);
}}
\end{center}
That is, pictorially we `erase' the names. Then,
because of the above identification, shifting a node through an $a-$ or $b-$ port is to do nothing: $T$ acts as the identity in the one particle sector. Since $H^2$ is the identity, $U^2$ also reduces to the identity in the one particle sector of the name invariant Hilbert space.  Pictorially, dropping the names $w, x, y, z$ in Fig. \ref{fig:UU}, we obtain Fig. \ref{fig:UU2}. The two last terms \emph{cancel} and the two first terms \emph{sum}. Two steps of the dynamics get us back to where we started. Therefore, with anonymous graphs and their span ${\cal H}'$ we are unable to express one of the simplest quantum walks.

{\em `Remote star' detection.} Now, if one insists on doing away with names, one may try to remedy this lack of descriptive power by providing relative position information in the graphs. After all, a `landmark' will be available in most practical situations, whether the laboratory walls or the `fixed stars'.  Attempting this ad hoc fix is instructive as it unveils an even more severe issue: depending on whether some arbitrarily remote landmark is present or not, the physics of the final state will be made to change in an instantaneous manner.

Let us illustrate this point through a gedanken experiment. We can model the presence of the landmark by introducing a new colour, bright green say, that is not affected by the dynamics. We will toggle this colour at some far away node $x$. Indeed, the calculation of Figure \ref{fig:UU} carries through when we consider circular graphs with an arbitrarily large number of nodes.\footnote{The same effect would be seen if we consider an infinite chain of nodes and edges.  The generalisation to infinite graphs is however beyond the scope of this paper, it would require the use of $C^*$-algebras for the definition of the inner product as was done for cellular automata in \cite{SchumacherWerner}.}

\begin{figure}[h!]
\includegraphics[width=\columnwidth]{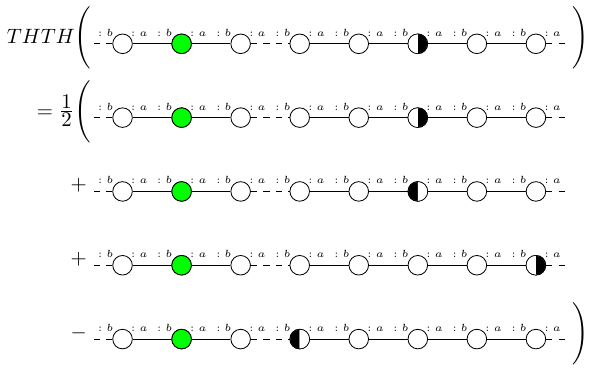}
\includegraphics[width=\columnwidth]{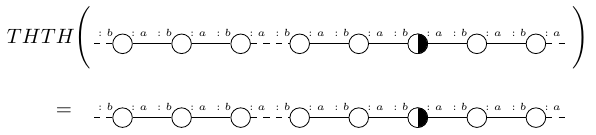}
\caption{\label{fig:star} Anonymous graphs can lead to instantaneous signalling. See the text for how this leads to a `remote star' detector. The graph segments shown here are understood as part of a large circular graph, or of an infinite chain of nodes. \\
{\em Top.} Local spin behaviour when star exists.\\
{\em Bottom.}  Local spin behaviour when star does not exist.}
\end{figure}

If the bright green star is present, it forbids the `unwanted identifications' between states. The correct behaviour of the dynamics is recovered, and a b--moving particle will be present in the state after application of $U$ twice, see Fig. \ref{fig:star} ({\em top}). When the bright green star is not present, the unwanted identifications cause the destructive interference of the b--moving degrees of freedom, see Fig. \ref{fig:star} ({\em bottom}). \emph{This change in spin particle behaviour is perfectly observable locally to the particle}. We thus obtain an arbitrarily remote `star' detector,  by observing the local statistics of the behaviour of spin particles. 

This is a type of instantaneous signalling, although we cautiously use this heavily connotative term in the following sense. In the parlance of quantum information operational theories, a far away agent $A$ can produce a bright green landmark, and this event is then signalled to an agent $B$ that is performing spin measurements after two time steps, regardless of the number of nodes and links between $A$ and $B$. $B$'s measurement does not require position or distance information, only the measurement capacity to discriminate the `spin' (a-- or b--moving) degree of freedom the particle. 

One might argue that the two agents at two specific nodes should be taken to be part of the state space, acting much like the landmarks considered previously. However, we can consider each node to be equipped with identical local measurement devices, that are triggered synchronously at $t=2$. Similarly, each node can be equipped with a quantum operation producing or not producing a landmark at $t=0$. Then, if any of the local observers observes a b--moving particle at $t=2$, they know for a fact that at least one landmark was produced at $t=0$, regardless of where this occurred. 

This instantaneous signalling is in contradiction with the basic principle of locality of interactions that underpins physics. If we hold on to this principle as being fundamental, then we must ask for formalisms that disallow instantaneous signalling -- even in a gedanken experiment. Taking the alternative route of employing both anonymous graphs and landmarks, it is likely that by positing `irremovable' landmarks on a case-by-case basis and appropriately sprinkled to break every symmetry of the application at hand, one might eventually remove unwanted identifications, forbidding instantaneous signalling. This ad hoc approach of checking how many `landmarks' would be sufficient for every dynamics and initial conditions of the theory at hand, could become a daunting task for a arbitrary graphs and arbitrary dynamics, potentially graph changing. The only way to be sure instantaneous signalling of this type is not present is to place distinguishable landmarks on all nodes of every graph---which is pretty much like naming the nodes to begin with, except that renaming-invariance goes unchecked. In our opinion, grappling with the theory in such a way would be like trying to hide the culprit under the carpet instead of recognising that it is a simple and elegant symmetry.  
 
This brings us back to the discussion of the previous section. Names are fiducial constructs that in the quantum theory are crucial for the purpose of aligning a superposition, and keeping the alignment through the dynamical evolution. Instead of trying to hide names under the carpet, we should explicitly allow for them, and posit invariance under renamings as a symmetry imposed on the theory. \VL{Conversely, having recognised invariance under renamings as a symmetry that is akin to a gauge invariance, we conclude that to avoid instantaneous signalling quantum theories built on graphs need to be based on a state space spanned by gauge variant states rather than on the gauge invariant space. Gauge invariance should be imposed later, at the level of the evolution and observables.}

\section{Formalisation: names versus colours}
\label{sec:labelledGraphs}

\VL{Moving to rigorous definitions, we clarify a simple but crucial point: the \emph{names} of graphs are a distinct, more primitive concept, than the \emph{colouring} of graphs. While it is true that when colouring a graph we may induce implicitly a renaming, it is conceptually imperative to separate the two. This is similar in the continuous to separating the procedure of laying coordinates on a manifold and that of adding physical fields that live on the manifold. We conclude the section with the definition of a Hilbert space built on graphs.}

We start by postulating a countable space ${\cal V}$ of possible names for the vertices of graphs. It suffices to think of them as the integers or `words' generated by a finite alphabet. We fix in addition $\pi$  to be a finite set of {\em ports} per node of the graph. We say that $f$ is a partial involution, meaning that if $f(x)$ is defined and equal to $y$, then $f(y)$ is defined and equal to $x$. 
 
\begin{definition}[Graph]\label{def:graphs}
Let $\pi$ be a finite set of {\em ports}. A {\em graph} $G$ is given by a pair $(V_G, \Gamma_G)$,  defined as
\begin{itemize}
\item[$\bullet$] A finite non-intersecting subset $V_G$ of ${\cal V}$. 
\item[$\bullet$] A partial involution $\Gamma_G$ from $V_G\port\pi$ to $V_G\port\pi$ thereby describing a set $E_G$:
$$\left\{\ \{u\port a,\Gamma_G(u\port a)\}\ |\ u\port a\in V_G\port\pi\ \right\}.$$ 
\end{itemize}
$V_G$ are the vertices, $E_G$ the edges and $\Gamma_G$ is the adjacency function of the graph. The set of all graphs is written ${\cal G}$.
\end{definition}
In usual definitions of graphs, edges connect nodes. Here they connect ports of nodes instead. There are two reasons for this choice. First, because it allows a node to distinguish its $\port a$ neighbour from its $\port b$ neighbour, which is necessary in order to express particle propagation as in the example of Sec. \ref{sec:signalling}. Second, because it allows to control the degree of the graph if necessary. For instance, in loop quantum gravity a node is dual to a tetrahedron and so the graph should be of degree less or equal to four.
\begin{definition}[Coloured Graph]\label{def:cgraphs}
A {\em coloured graph} $(G,C_G)$ is a graph $G$ with the addition of $C_G \equiv (\sigma_G, \delta_G)$:
\begin{itemize}
\item[$\bullet$] A partial function $\sigma_G$ from $V_G$ to a set $\Sigma$. 
\item[$\bullet$] A partial function $\delta_G$ from $E_G$ to a set $\Delta$.
\end{itemize}
The sets $\Sigma$ and $\Delta$ are the internal states, a.k.a node/vertex colours and links/edge state colours of the theory. The set of coloured graphs is written ${\cal G}_C$. 
\end{definition}

With these two definitions, we see that \emph{assigning names} to the vertices of a graph is but a mathematical tool for describing them, similar to a coordinate system on a manifold. 
The \emph{colouring} on the other hand, serves the purpose of encoding a `field'---depending on the application at hand-- and comes on top of the naming of nodes. 
\VL{For instance, as discussed in Section \ref{sec:QG} a colouring of graphs can code for quantum spacetime geometry.}
For matter, quite often fermionic fields are modelled as living on the nodes, whereas bosonic fields are modelled as living on the edges \cite{ArrighiQED}. In summary, names and colours should not be confused. \VL{Names play the role of coordinates on the graph and are essential for describing the graph itself. Colours play the role of the physical field living on top of the graph.} In particular, it can of course happen that a field (colouring) can take the same value $\sigma(u)=\sigma(v)$ at two different places $u\neq v$, which is not allowed for names.  
These observations suggest that approaches dealing with superpositions of manifolds that label points by the values of auxiliary fields, such as \cite{hardy2020implementation}, may suffer from the instantaneous signalling of the previous section. Indeed, in this path integral approach diffeomorphisms are allowed to act independently on each quantum branch, which suggests that the underlying \VL{kinematical }state space considered is spanned by name invariant states.

For the sake of clarifying the argument of Section \ref{sec:signalling}, let us now define anonymous graphs. This is done by considering equivalence classes of named graphs modulo isomorphism:
\begin{definition}[Anonymous graphs]\label{def:pointedmodulo}
The graphs $G, H\in {\cal G}$ are said to be isomorphic if and only if there exists an injective function $R(.)$ from ${\cal V}$ to ${\cal V}$ which is such that $V_H=R(V_G)$ and for all $u,v\in V_G, a,b\in\pi$, $R(v)\port b=\Gamma_H(R(u)\port a)$ if and only if $v\port b=\Gamma_G(u\port a)$. Additionally, if the graphs are coloured  $\sigma_H\circ R=\sigma_G$ and for all $u,v\in V_G, a,b\in\pi$, $\delta_H(\{R(u)\port a,R(v)\port b\})=\delta_G(\{u\port a,v\port b\})$. We then write $G\sim H$. Consider $G\in{\cal G}$. The {\em anonymous graph} $\widetilde{G}$ is the equivalence class of $G$ with respect to the equivalence relation $\sim$. Similarly if $G\in{\cal G}_C$. The space of anonymous graphs is denoted $\tilde{\mathcal{G}}$.
\end{definition}
Notice how, in the above definitions, anonymous graphs arise from graphs and not the reverse way round---a common misconception in Physics. In fact, we are not aware of a way to mathematically define a generic anonymous graph first, without any reference to a (named) graph, and then assign names to its nodes. 
\VL{Graphs, a concept that presupposes naming the nodes, are the primitive mathematical notion, from which anonymous graphs can be derived.}


We argued in the previous section that quantum superpositions of graphs can and must be defined as the span of named graphs. Therefore, we construct a state space based solely based on Definition \ref{def:graphs}:
\begin{definition}[Superpositions of graphs]~\label{def:HCfbis} We define ${\cal H}$ the Hilbert space of graphs, as that having $\{\ket{G}\}_{G\in{\cal G}_C}$ as its canonical orthonormal basis.
\end{definition}
This definition captures the main point of this paper: ${\cal G}_C$ is the set of coloured graphs of Definition \ref{def:cgraphs}, that inherits the names from the Definition \ref{def:graphs} for graphs on which it is based. We have argued that it is this state space that should be the basis of a theory that manipulates superpositions of graphs, rather than a state space based on the name invariant space $\tilde{{\cal G}}$ of Definition \ref{def:pointedmodulo}. Thus, we posit:
\begin{postulate}
Physically relevant quantum superpositions of graphs are elements of ${\cal H}$. 
\end{postulate}
As usual in quantum theory, states can either be `state vectors' (pure states), i.e. unit vectors $\ket{\psi}$ in ${\cal H}$, or `density matrices' (possibly mixed states), i.e. unit trace non-negative operators $\rho$ over ${\cal H}$. Evolutions can be prescribed by unitary operators $U$ over  ${\cal H}$, taking $\ket{\psi}$ into $U\ket{\psi}$, or alternatively $\rho$ into $U\rho U^\dagger$.

\section{Renamings, observables and evolutions}
\label{sec:observables}

We now proceed to treat renamings as a symmetry group of our quantum state space. In the above section, the definition of `renamings' (graph isomorphism) was inlined inside Definition \ref{def:pointedmodulo} of anonymous graphs. Having argued we should not work at the level of name invariant graphs, but at the level of (name variant) graphs, renaming invariance remains to be enforced. First, we define renamings as a standalone notion acting on the state space $\cal{H}$ of graphs:
\begin{definition}[Renaming]\label{def:graph renaming}
Consider $R$ an injective function from ${\cal V}$ to ${\cal V}$.
Renamings act on elements of ${\cal G}$ by renaming each vertex, and are extended to act on ${\cal H}$ by linearity, i.e. $R\ket{G}=\ket{RG}$ and $\bra{G}R^\dagger=\bra{RG}$.
\end{definition}

\medskip

\noindent {\em Observables.} Physically relevant observables must be name-invariant, so that probabilities or expected values given by the Born rule are unaltered under renaming. Thus, we must demand that {\em global} observables satisfy 
$$
\tr\left(O R\ket{G}\!\bra{G}R^\dagger \right)=\tr\left(R^\dagger O R \ket{G}\!\bra{G} \right)=\tr\left(O \ket{G}\!\bra{G} \right)
$$
which follows for all $G$ if and only if $[R,O]=0$ where $[,]$ is the commutator of operators on $\mathcal{H}$. An example of a valid global observable is the total number of links and nodes of the graph. An example of an invalid global observable is $\sum_{u\in V(G)} \ket{G}\bra{G}$ the projector upon the subspace of those graphs which contain a node named $u$.

We call {\em $1$-point} observable a function from ${\cal V}$ to operators over ${\cal H}$, i.e. a map $u\mapsto O(u)$. This is basically an observable which is parameterised by $u$, typically because it is centred around node $u$. The {\em local observables at $u$}, morally of the form $O(u)=L\otimes \bigotimes_{v\neq u} \mathds{1}$, are a typical class of $1$-point observables. But this can also include larger radius observables. E.g.
we may be interested in knowing `whether a particle is present on the neighbour of $u$ along port $\ \port a$'. Renaming the entire graph, and node $u$ into node $v$ in particular, does not alter this fact, so long as we now ask `whether a particle is present on the neighbour of $v$ along port $\ \port a$' instead. Thus, for the sake of allowing for $1$-point observables we need to refine our previous requirement as follows:
$$ 
\tr\left(O(u)\ket{G}\!\bra{G}\right)=\tr\left(O(R(u))R\ket{G}\!\bra{G}R^\dagger\right)
$$
i.e. now the Born rule does depend upon $u$, but is some mild manner.
We reach the following definition:
\begin{definition}[Renaming invariance]
An operator
$O(u)$ is said to be renaming invariant if and only if for all $u\in {\cal V}$, for all $G\in \cal G$ and for all renamings $R$, 
$O(R(u))R=R O(u)$.
\end{definition}
This generalizes to $n$-point observables: $u$ can be understood as a list of nodes $u_1,\ldots,u_n$ in the above definition. When the list is empty, we recover renaming invariance for global observables.\\
We insist that this is a non-trivial criterion upon the function $u\mapsto O(u)$. Indeed, with the above definition we are effectively forbidding `observing' the `name-dependent quantities'. For instance, say that nodes are numbered by integers. Measuring `the quantity that corresponds to the name of the neighbour of $u$ along port $\ \port a$' is not a valid $1$-point observable, because it is not renaming invariant. Measuring `when $u$ is even whether a particle is present at $u$, and when $u$ is odd whether no particle is present at $u$' is not a valid $1$-point observable either, because it is not renaming invariant.\\
An interesting example of a valid $1$-point observable is that which reads out the ratio between the number of second neighbours, and the number of first neighbours, of a node $u$---this is often thought of as a discrete analogue of a Ricci scalar curvature for graphs.\\
The ability to define $n$-point observables, whilst still being able to identify the renaming invariant ones, is no doubt another strong point in favour of formalisation proposed. With anonymous graphs instead, we would be unable to even express such quantities.


\medskip

\noindent {\em Evolutions.} Physically relevant evolutions need be insensitive to the names of the vertices. Thus, we must demand that any global evolution $U$ be renaming invariant in the following sense: for any renaming $R$, we must have that $UR = RU$. Examples of valid evolutions were given in Sec. \ref{sec:signalling}. Recall that amongst them, the global evolution $H$ decomposed into local $\textrm{Hadamard}$ evolutions. In order to speak of each of these local evolutions individually it may again convenient be convenient to use a $1$-point evolution $\textrm{Hadamard}(u)$. It verifies $\textrm{Hadamard}(R(u))R = R\,\textrm{Hadamard}(u)$. Another use of $1$-point evolutions is in order to model agency and study causality, i.e. agent Alice causing operation $A(u)$ at time $t$ and position $u$, having effect $B(u)=UA(u)U^\dagger$ at time $t+1$ in the neighbourhood of $u$ under the global evolution $U$. We, thus, posit:
\begin{postulate}
Physically relevant observables over quantum superpositions of graphs are renaming invariant operators over ${\cal H}$. 
\end{postulate}
A keen reader will ask whether this last postulate is compatible with the possibility that an evolution may create/destruct nodes. There is indeed a subtlety here. One of the authors \cite{ArrighiRCGD} has shown that in the classical, reversible setting, and when using straightforward naming conventions for the nodes, renaming invariance enforces node preservation. However, node creation/destruction becomes possible again when we adopt slightly more elaborate naming schemes, as shown by two of the authors in \cite{ArrighiDestruction,ArrighiQNT}.

\section{Relevance for Quantum Gravity}
\label{sec:QG}
Diffeomorphism invariance is the central symmetry underlying general relativity. Its physical content is to ensure that predictions of the theory do not depend on the choice of a coordinate system. This is because (smooth) changes of coordinates are in a one--to--one correspondence with diffeomorphisms. Because of this mathematical equivalence, changes of coordinates are called `passive diffeomorphisms' while the more abstract notion of a diffeomorphism as a map on the manifold is called an `active diffeomorphism'. Similarly, graph renamings  can be thought of as the `passive' way of describing graph isomorphisms. 

 Diffeomorphisms are a primitive concept defined already at the pre-geometrical level of a manifold, before we consider the metric tensor that describes the spacetime or matter fields. In this work we identified renamings of graphs as a discrete analogue to coordinate changes on a manifold. The, now discrete,  pre-geometrical space is the graph.  The assignment of names to nodes is the assignment of `coordinates' to the set of points of this space, the graph nodes. Links are to be understood as an adjacency relation in the topological, pre-geometrical, sense.

Let us discuss our findings in the context of Loop Quantum Gravity (LQG). In this well developed tentative theory for quantum gravity, a central result is that the kinematical state space decomposes in Hilbert spaces each corresponding to a graph, spanned by a certain basis of colourings of this graph called the spin--network states. The graph can be understood as dual to a `triangulation' of three dimensional space with quantum tetrahedra (see for instance\cite{PhysRevD.83.044035}). This is a quantum geometry in the sense that  geometrical observables such as areas and angles do not commute and thus are undetermined satisfying uncertainty relations.

In the literature, we find two ways to enforce spatial diffeomorphism invariance at the quantum level in this program. One way  \cite{rovelli_vidotto_2014} is to embed the graphs in a manifold, such that links become curves on the manifold, and then take the equivalence class resulting from acting with spatial diffeomorphisms on the embedded graph. This is a tedious procedure leading to a number of complications such as the creation of knots. Another strategy  \cite{thiemann_2007}, at odds with the former approach, is to work directly at the level of non-embedded graphs, also called `abstract graphs'. The latter method is claimed to completely do away with the need to impose the (spatial) diffeomorphism invariance constraint at the quantum level, as there is no embedding manifold to begin with. This is a minimalist and much simpler point of view.

Our analysis suggests that both points of view are partly correct and partly misplaced. In the former approach where the graph is embedded in a manifold, it seems superfluous to consider an additional continuous background space if the theory is to be based on graphs. Graphs already serve the role of a natively discrete topological ambient space on which fields can then be defined. In the latter approach that employs `abstract' or non--embedded graphs, as we do here, it is misplaced to claim that the invariance under changes of coordinates, a central symmetry of the classical theory, has disappeared altogether on the grounds of an embedding manifold not being present. Whilst the embedding continuous topological space is appropriately dropped, on the discrete topological structure that remains---the graph--- there still remains the possibility to rename the nodes of the graph. Thus, in the context of LQG, we may introduce renaming invariance as implementing spatial diffeomorphism invariance at the quantum level.  

The above relate to describing a superposition of spacetimes as follows. We have seen that the names used in the definition of graphs play a crucial role in aligning superpositions of graphs. In LQG, a superposition of (quantum) spacetimes, is represented as a superposition of appropriately colored graphs. The simplest case conceptually is to consider a superposition of two coloured graph states$ \ket{G}$ and $\ket{H}$ each corresponding to  a superposition of distinct semiclassical  spatial geometries, a superposition of two `wavepackets' of geometry (see for instance \cite{phdM} for an introduction) each peaked on the 3-geometry of a spacelike surface.  Consider then the state 
\begin{equation}
\ket{G}+\ket{H}
\end{equation}
and let us momentarily consider embedding the two graphs $G$ and $H$ in a three dimensional Riemannian manifold. We consider a node $u$ which exists both in $G$ and $H$.  The graph embedding can be defined in a common coordinate system $(x^\alpha)$. There are  two different metric fields $g(x^\alpha)$ and $h(x^\alpha)$ defined on the manifold, on which the state of each graph is correspondingly peaked. In particular, the colouring of the node $u$ will be different in the two graphs . In the manifold, the node has coordinates $x^\alpha_u$. A diffeomorphism $\phi$ will change these coordinates to $\phi^* x^\alpha_u$. This is simply a change of name for the node. Of course, diffeomorphisms would give a continuous range of possible names, making the use of real numbers necessary. Working in the discrete, a countable range of names is enough. The key point is that the induced renaming is the same on both $G$ and $H$. From Definition 5, it acts on all branches of the superposition, as in
\begin{equation}
R \ket{G}+ R \ket{H}.
\end{equation}
This the non-trivial content of renaming invariance we have seen above in Section \ref{sec:signalling}. The renaming $R$, acts as a a sort of `quantum diffeomorphism', because it is acting on a superposition of graphs. It preserves the `alignment' of the superposition, regardless of whether a colouring is present or not. Having now control of this, we may proceed to colour the graph appropriately with discrete fields that admit a physical interpretation, as above, and proceed to describe interesting physics such as a superposition of spacetimes. That is, in the example above, it will remain the case under arbitrary renamings that the renamed node $u$ will have the same colouring in the two graph states, thus, also the two values of the geometrical data captured in $G$ and $H$ at that node will remained superposed but aligned. We stress once again that embedding the graphs in an ambient manifold is a superfluous procedure, that was employed here for the purpose of demonstration. It is sufficient to work at the level of graphs, recognising that they carry names on their nodes by their very definition. Then, renaming invariance naturally arises as a native discrete analogue of diffeomorphism invariance, that can be carried through at the quantum level.

\bigskip

\section{Conclusions}

{\em Summary of contributions.} We provided a robust notion of quantum superposition of graphs, to serve as the state space for various applications, ranging from distributed models of Quantum Computing, the `quantum internet', to Quantum Gravity. 
 
The main difficulty lied in the treatment of names. While node names are part of the definition of graphs they need to be done away with. We have shown that getting rid of them by working at the level of `anonymous' graphs is problematic as it leads to instantaneous signalling. We pointed out that the underlying reason is that names play an essential role in keeping quantum superpositions of graphs `aligned' with respect to one another. 
 
We then proceeded to define renamings as a symmetry of the quantum state space and postulated that observables and evolutions should satisfy renaming invariance, i.e. they must commute with renamings. Conveniently, the way we defined renaming-invariant still allowed us to talk about observables centered at node $u$, which would not be possible with anonymous graphs.

\VL{We pointed out that renamings on graphs are the native discrete analogue of passive diffeomorphisms on a manifold---both are relabelling the points of a topological space. In this sense, we followed here the standard prescription of General Relativity: use coordinates to define the physical situation being studied, and then demand that statements of physical relevance be invariant under changes of coordinates. Having furthermore straightforwardly extended graph renamings at the quantum level, we suggest them as a strategy for implementing discrete diffeomorphism invariance at the quantum level.}

{\em Some perspectives.} In the continuous setting, passive diffeomorphisms (aka differentiable changes of coordinates) and active diffeomorphisms (aka differentiable displacements of the points in space) coincide: they are but two sides of the same coin. In the discrete this is by no means obvious: whilst the former amount to graph renamings as we have seen, the latter have often been interpreted as changing the shape of graph, to other triangulations of the same underlying manifold. Discrete analogues to active diffeomorphism, and their generated Hamiltonian constraints were studied e.g. in \cite{dittrich2009diffeomorphism,hohn2015canonical,hohn2014quantization}. The invariance that were found were not exact, except in flat space. Which of the passive or active picture is physically relevant to the discrete? Can build on the exact renaming invariance described here, as well as the passive/active analogy of the continuum, to work towards exact retriangulation invariance?     

\VL{Whilst this paper focused on graphs (understood as a basis for 3--dimensional quantum space in LQG), the same line of argument could be carried through on `higher dimensional' graphs, e.g. 2--dimensional cellular complexes which correspond to a 4--dimensional quantum spacetime in LQG. Renaming invariance will then serve as a discrete analogue of full, 4--dimensional diffeomorphism invariance.}

\VL{Having a robust notion of quantum superposition of graphs allows us to define quantum observables even at the pre-geometric level, before any spacetime geometry emerges or a matter field is considered, implying that information can be stored at the pre-geometric level. This potentially vast storage space could be holding the key to solving the black hole information loss paradox, as suggested in \cite{Perez:2014xca}.} Quantum superpositions of graphs may also provide a discrete, reference frame-independent formalism for the recent continuous theories of quantum coordinate systems \cite{Hardy:2019cef} and quantum reference frames \cite{Castro-Ruiz:2019nnl,Giacomini:2017zju}. A quantum computing perspective is to encode indefinite causal orders \cite{OreshkovQuantumOrder} directly as quantum superpositions of directed acyclic graphs.

\paragraph{Acknowledgments.} The authors thank \c{C}aslav Brukner, Lucien Hardy, Aristotelis Panagiotopoulos, Alejandro Perez and Carlo Rovelli, for insights and discussions on this work. This project/publication was made possible through the support of the ID\# 62312 grant from the John Templeton Foundation, as part of the \href{https://www.templeton.org/grant/the-quantum-information-structure-of-spacetime-qiss-second-phase}{‘The Quantum Information Structure of Spacetime’ Project (QISS) }. The opinions expressed in this project/publication are those of the author(s) and do not necessarily reflect the views of the John Templeton Foundation. It was also supported by the PEPR integrated project EPiQ ANR-22-PETQ-0007 part of Plan France 2030. 

\bibliographystyle{plain}
\bibliography{biblio2,gravQSwitch,discreteness,pablo,biblio}

\end{document}